\documentclass[reprint,amsmath,amssymb,aps,noeprint,floatfix, nolinenumbers,longbibliography]{revtex4-2}

\usepackage[english]{babel}
\usepackage[utf8]{inputenc}
\usepackage[colorinlistoftodos, color=green!40, prependcaption]{todonotes}
\usepackage{float}
\usepackage{graphicx}
\usepackage{yhmath}
\usepackage{subfigure}
\usepackage{mathdots}
\usepackage{MnSymbol}
\usepackage{dsfont}
\usepackage{soul}
\usepackage[normalem]{ulem}
\usepackage{hyperref}
\usepackage{bm}
\usepackage{multirow}

\hypersetup{
    unicode=true, 
    plainpages=false,
    colorlinks=true,
    linkcolor=blue,
    citecolor=blue,
    filecolor=blue,
    urlcolor=blue
}

\usepackage{mathtools}

\DeclarePairedDelimiterX\braket[2]{\langle}{\rangle}{#1\,\delimsize\vert\,\mathopen{}#2}

\newcommand{\md}{\mathrm{d}}

\begin{document}
\title{Complex Vector Gain-Based Annealer for Minimizing  XY Hamiltonians}

\author{James S. Cummins and Natalia G. Berloff}
\email[correspondence address: ]{N.G.Berloff@damtp.cam.ac.uk}
\affiliation{Department of Applied Mathematics and Theoretical Physics, University of Cambridge, Wilberforce Road, Cambridge CB3 0WA, United Kingdom}

\date{\today}

\begin{abstract}
This paper presents the Complex Vector Gain-Based Annealer (CoVeGA), an analog computing platform designed to overcome energy barriers in XY Hamiltonians through a higher-dimensional representation.  Traditional gain-based solvers utilizing optical or photonic hardware typically represent each XY spin with a single complex field. These solvers often struggle with large energy barriers in complex landscapes, leading to relaxation into excited states. CoVeGA addresses these limitations by employing two complex fields to represent each XY spin and dynamically evolving the energy landscape through time-dependent annealing. Operating in a higher-dimensional space, CoVeGA bridges energy barriers in this expanded space during the continuous phase evolution, thus avoiding entrapment in local minima. We introduce several graph structures that pose challenges for XY minimization and use them to benchmark CoVeGA against single-dimension XY solvers, highlighting the benefits of higher-dimensional operation.
\end{abstract}

\maketitle

\section{Introduction}

The growing complexity and sheer scale of modern scientific and industrial computing tasks are pushing us to look beyond traditional von Neumann architectures for solutions to hard optimization problems. These architectures, which dominate computing today, rely on a clear separation between memory and processing and execute tasks in a step-by-step manner. While traditional computing has been reliable for decades, it’s starting to fall behind when it comes to the needs of today’s specialized applications—particularly those that demand high speed, energy efficiency, and scalability. As areas like machine learning, big data analysis, and real-time processing continue to grow, the limitations of the von Neumann approach have become more evident, creating a bottleneck that is increasingly difficult to overcome. This is where analog systems come in, offering a tailored approach to specific types of computing tasks, bypassing the constraints of conventional architectures.

Physics-inspired analog machines have been proposed using various platforms such as superconducting qubits \cite{johnson2011quantum, denchev2016computational, arute2019quantum}, optical parametric oscillators \cite{mcmahon2016fully, inagaki2016coherent, yamamoto2017coherent, honjo2021100}, memristors \cite{cai2020power}, lasers \cite{babaeian2019single, pal2020rapid, parto2020realizing}, photonic systems \cite{pierangeli2019large, roques2020heuristic}, trapped ions \cite{kim2010quantum}, polariton condensates \cite{berloff2017realizing, kalinin2020polaritonic}, photon condensates \cite{vretenar2021controllable}, and surface acoustic waves \cite{litvinenko202350}. These specialized physical machines minimize programmable spin Hamiltonians, where the couplings between spins -- given by the interaction matrix $\mathbf{J}$ -- are designed such that the global minimum corresponds to the optimal solution of a combinatorial optimization problem. Problems like number partitioning, traveling salesman, graph coloring, spin glass systems, knapsack problem, binary linear programming, graph partitioning, and Max-Cut can be mapped to spin Hamiltonians \cite{lucas2014ising, yamaoka201520k, tanahashi2019application}. Further applications include machine learning \cite{momeni2024training}, financial markets \cite{gilli2019numerical}, and portfolio optimization \cite{wang2024efficient}. Most proposed physical analog machines use one-dimensional (`hard' or `soft') spins as variables in the discrete binary Ising Hamiltonian $H_{\rm I} = - \sum_{i, j}^N J_{ij} s_i s_j$ with $s_i = \pm 1$, or the continuous XY Hamiltonian $H_{\rm XY} = - \sum_{i, j}^N J_{ij} {\bf s}_i \cdot {\bf s}_j$, where ${\bf s}_i = ( \cos \theta_i, \sin \theta_i )$, $\theta_i \in [0, 2 \pi)$, and $N$ is the number of spins.

The XY model describes a system of spins constrained to rotate within a plane, each possessing a continuous degree of freedom characterized by a phase $\theta_{i}$ relative to a fixed axis. Interactions between phases give rise to a rich array of dynamics, including vortices and Berezinski-Kosterlitz-Thouless transitions \cite{kosterlitz2018ordering}. Originally developed for statistical mechanics and condensed matter physics, the XY model has found extensive applications in various physical systems where rotational symmetry is essential. Notably, it applies to superfluidity and superconductivity \cite{minnhagen1987two}, cosmology \cite{zurek1995cosmological}, nematic liquid crystals \cite{de1993physics}, magnetic nanoparticle ensembles \cite{gallina2020disorder}, protein folding \cite{dill2008protein}, and phase retrieval problems \cite{candes2015phase}.

Gain-based minimizers utilize soft-spin bifurcation dynamics via Andronov-Hopf bifurcations to minimize spin Hamiltonians \cite{syed2023physics}. The enhanced dimensionality offered by soft-spin models reduces energy barriers present in classical hard-spin Hamiltonians by representing fixed spin amplitudes as continuous variables \cite{cummins2023classical}. We recently proposed the Vector Ising Spin Annealer (VISA)  as an Ising minimization model capable of overcoming obstacles in solving combinatorial optimization problems \cite{cummins2024vector}. By employing three soft modes to represent the vector components of an Ising spin, VISA bridges minima separated by significant energy barriers in complex energy landscapes. VISA uses real-valued soft spins, which can, for example, represent the optical parametric oscillator quadrature in coherent Ising machines.

In this paper, we propose to use multiple vector components to represent the spins in networks of complex-valued fields $\psi_i$, which are known to minimize XY and Ising Hamiltonians in gain-based systems \cite{kalinin2018global}. The Stuart-Landau equation that governs the dynamics of one-dimensional complex oscillators $\psi_{i}$ in gain-based networks is given by
\begin{equation} \label{Gain-Diss Eq}
    \Dot{\psi}_{i} = \left( \gamma_{i} - |\psi_{i}|^2 \right) \psi_{i} + \alpha \sum_{j} J_{ij} \psi_{j},
\end{equation}
where we introduced a regulation parameter $\alpha > 0$ and the effective gain rate $\gamma_{i}$ is individually dynamically adjusted through feedback mechanism
\begin{equation} \label{Feedback Eq}
    \Dot{\gamma}_{i} = \varepsilon \left(1 - |\psi_{i}|^2 \right),
\end{equation}
which can be implemented via optical delay lines for laser systems, or by spatial light modulators for polariton condensates \cite{kalinin2020polaritonic}; see also Appendix \ref{Section:cGLE}. The dynamical equation (\ref{Gain-Diss Eq}) can be written as $\Dot{\psi}_{i} = - \partial H / \partial \psi_{i}^{*}$, which describes the process of gradient descent to the minima of the loss function
\begin{equation} \label{Hamiltonian Eq}
    H = \frac{1}{2} \sum_{i = 1}^N \left( \gamma_{i} - |\psi_{i}|^2 \right)^2 - \frac{\alpha}{2} \sum_{i,j}^N J_{ij} \left( \psi_{i} \psi_{j}^{*} + \psi_{i}^{*} \psi_{j} \right),
\end{equation}
where $\dot{\gamma}_i = 0$ for all $i$ at the threshold steady state. By representing each oscillator in its polar form as $\psi_{i} = r_{i} \exp (i \theta_{i})$, Eq.~(\ref{Gain-Diss Eq}) can be decomposed into real and imaginary parts to get equations of the time evolution of the amplitude $r_{i}$ and the phase $\theta_{i}$ as
\begin{align}
    \dot{r}_{i} & = \gamma_{i} r_{i} - r_{i}^{3} + \alpha \sum_{j} J_{ij} r_{j} \cos \left( \theta_{i} - \theta_{j} \right), \label{Density Eq} \\
    \dot{\theta}_{i} & = - \alpha \sum_{j} J_{ij} \frac{r_{j}}{r_{i}} \sin \left( \theta_{i} - \theta_{j} \right). \label{Phase Eq}
\end{align}
Starting from below the steady state threshold in the vacuum state $r_{i} = 0$, all oscillators are pumped equally. Then, depending on the structure of $\mathbf{J}$, nonzero amplitudes emerge at different rates for each oscillator as the pumping intensity increases. The feedback mechanism of Eq.~(\ref{Feedback Eq}) adjusts each oscillator so that they all reach equal amplitudes at the steady state threshold.
 
Only under the condition of equal amplitudes at the steady state will Eq.~(\ref{Phase Eq}) reach the minimum of the XY Hamiltonian. The sum of the steady states of Eq.~(\ref{Density Eq}) gives $N = \sum_{i = 1}^{N} \gamma_{i} +\alpha / 2 \sum_{i, j} J_{ij} \cos \left( \theta_{i} - \theta_{j} \right)$, so the global minimum of the XY model corresponds to the smallest effective injection $\sum_{i} \gamma_{i}$. Close to the threshold, Eq.~(\ref{Phase Eq}) becomes fully analogous to the Kuramoto model
\begin{equation} \label{Kuramoto Equation}
    \Dot{\theta}_{i} = - \alpha \sum_{j=1}^{N} J_{ij} \sin \left( \theta_{i} - \theta_{j} \right).
\end{equation}
Equation (\ref{Gain-Diss Eq}) can be adapted to minimize Ising Hamiltonians by restricting the state space of the phase, which we detail in Appendix \ref{Section:XY to Ising}.

Gain-based systems described by Eq.~(\ref{Gain-Diss Eq}) can still settle in local minima during amplitude bifurcation, which limits the probability of finding the global minimum. To combat this, we introduce the complex vector gain-based annealer (CoVeGA) that exploits the advantages of extended spatial dimensions. In this model, continuous XY spins are represented as complex-valued vectors in two-dimensional vector space ${\bf \Psi}_i \in \mathbb{C}^{2}$. The increased dimension over typical one-dimensional spins allows to effectively overcome the barriers between minima.

This paper proposes a new approach to minimizing XY Hamiltonians that utilizes the ultra-fast energy-efficient architecture of photonics-based analog machines. In Section \ref{Section:CVGBA}, we formalize CoVeGA, and provide expressions for each term in its composite Hamiltonian. In Section \ref{Section:Graphs}, we use the Kuramoto model to investigate the difficulty of various XY minimization problems. This allows us to identify suitably hard benchmark problems to test CoVeGA and existing XY minimization algorithms, as recovering the global minimum of these problems is nontrivial. Lastly, Section \ref{Section:Results} compares the dynamics of CoVeGA to the one-dimensional Stuart-Landau network as well as other continuous-variable methods such as spin-vector Langevin and Kuramoto models. We contrast these methods by finding ground and excited state probabilities and illustrating the distribution of recovered states.

\section{Complex Vector Gain-Based Annealer}
\label{Section:CVGBA}

The CoVeGA model operates through a system of $N$ two-dimensional complex-field vectors ${\bf \Psi}_i =(\psi_i^{(1)} = r_{i}^{(1)} e^{i \theta_{i}^{(1)}}, \psi_i^{(2)} = r_{i}^{(2)} e^{i \theta_{i}^{(2)}})$. It utilizes annealing, symmetry-breaking bifurcation, gradient descent, and mode selection to drive the system to the global minimum. The Hamiltonian is the sum of three terms $H = H_1 + \alpha H_2 + H_3$, where
\begin{align}
    H_1 & = \frac{1}{2} \sum_{i = 1}^{N} \left( \gamma_i(t) - || {\bf \Psi}_i ||_{2}^{2} \right)^2, \label{H1} \\
    H_2 & = - \frac{1}{2} \sum_{i, j}^{N} J_{ij} \left( {\bf \Psi}_i \cdot {\bf \Psi}_{j}^{*} + {\bf \Psi}_{i}^{*} \cdot {\bf \Psi}_{j} \right), \label{H2} \\
    H_3 & = - P(t) \sum_{i, j = 1}^N \left( {\bf \Psi}_{j} \odot {\bf \Psi}_{i}^{*} \right) \cdot \left[ {\bf Q} ({\bf \Psi}_{j}^{*} \odot {\bf \Psi}_{i}) \right]. \label{H3}
\end{align}
Here, $\odot$ indicates element-wise multiplication, and ${\bf Q}$ is a $2 \times 2$ permutation matrix given by ${\bf Q} = \bigl( \begin{smallmatrix} 0 & 1 \\ 1 & 0 \end{smallmatrix}\bigr)$. As the effective gain $\gamma_i (t)$ increases with time $t$ from negative (effective losses) to positive values, $H_1$ anneals between a convex function with minimum at $|| {\bf \Psi}_{i} ||_{2}^{2} = 0$ for all $i$, to nonzero amplitudes. Writing the complex vectors in $H_2$ using their polar coordinates  gives
\begin{equation} \label{H2 Expansion}
    H_2 = - \sum_{i, j} J_{ij} \left( r_{i}^{(1)} r_{j}^{(1)} \cos \theta_{ij}^{(1)} + r_{i}^{(2)} r_{j}^{(2)} \cos \theta_{ij}^{(2)} \right),
\end{equation}
where we define $\theta_{ij}^{(k)} \equiv \theta_{i}^{(k)} - \theta_{j}^{(k)}$. Equation (\ref{H2 Expansion}) is analogous to the second term on the right-hand side of Eq.~(\ref{Hamiltonian Eq}), but now we have two terms corresponding to the two dimensions in the complex vector space that CoVeGA operates in. The effective gain is subject to the feedback governed by $\Dot{\gamma}_i = \varepsilon (1 - || {\bf \Psi}_{i} ||_{2}^{2})$. This drives the effective gain rates from below and   brings the amplitudes to $1$ at the steady state. Finally, $H_{3}$ is a penalty term with time-dependent magnitude $P(t)$ that enforces agreement of the phase differences in each dimension. This $4$-local penalty term, expressed in Eq.~(\ref{H3}), can be expanded into its constituent complex fields, which transforms $H_3$ to
\begin{equation} \label{Expanded H3}
    H_3 = - P(t) \sum_{i,j} \psi_i^{(1)} \psi_i^{(2)*} \psi_j^{(1)*} \psi_j^{(2)} + \text{c.c.}
\end{equation}
When the complex oscillators have equal unit-valued amplitudes, $|| {\bf \Psi}_{i} ||_{2}^{2} = 1$ for all $i$,  $H_3$ becomes
\begin{equation}
    H_3 = - 2 P(t) \sum_{i,j} \cos \left( \theta_{ij}^{(1)} - \theta_{ij}^{(2)} \right),
\end{equation}
which is minimized when $\theta_{ij}^{(1)} =  \theta_{ij}^{(2)}$ for all pairs $(i, j)$. As $P(t)$ increases from $P(0) = 0$ to sufficiently large $P(T) > 0$ at $t = T$, the phase differences $\theta_{ij}^{(1)}$ and $\theta_{ij}^{(2)}$ become equal. Therefore, up to a global phase offset, $\theta_{i}^{(1)} = \theta_{i}^{(2)}$. At the end of the annealing protocol, the target XY Hamiltonian $H_{\rm XY}$ is minimized. We evolve the CoVeGA Hamiltonian $H$ using gradient descent while simultaneously annealing parameters $\gamma_{i}$ and $P(t)$. Therefore, for each oscillator $i$, the governing equation $\Dot{{\bf \Psi}}_{i} = - \nabla_{i} H$ is given by
\begin{equation} \label{Governing Equation}
    \begin{split}
    \Dot{{\bf \Psi}}_{i} & = {\bf \Psi}_{i} \left( \gamma_i (t) - || {\bf \Psi}_{i} ||_{2}^{2} \right) + \alpha \sum_{j = 1}^{N} J_{ij} {\bf \Psi}_{j} \\
    & + P(t) \sum_{j = 1}^{N} {\bf \Psi}_{j} \odot \left[ {\bf Q} ({\bf \Psi}_{j}^{*} \odot {\bf \Psi}_{i}) \right].
    \end{split}
\end{equation}
The operation of CoVeGA, therefore, relies on the gradient descent of an annealed energy landscape. The Hamiltonian $H$ is $4$-local due to the $H_3$ penalty term. While a $4$-local Ising Hamiltonian can be relaxed to a quadratic function without loss of generality \cite{boros2002pseudo}, subject to an overhead corresponding to the introduction of $N^{\lceil k/2 \rceil}$ auxiliary variables, this is not the case for XY Hamiltonians. In general, a $4$-local XY Hamiltonian can not be directly mapped to a $2$-local Hamiltonian. However, some optical hardware can directly encode such high-order interactions for XY systems \cite{stroev2021discrete}.

In the next section, we seek suitable graph structures for benchmarking CoVeGA and alternate algorithms for XY Hamiltonian minimization. For benchmarking, we choose graphs with analytically tractable yet nontrivial ground states and energies and whose basins of attraction are small in volume. In this way, simple gradient-based algorithms are expected to falter in recovering the global minimum. Moreover, we seek technologically feasible graphs on analog hardware \cite{ohadi2016nontrivial, strinati2021all, ayoub2021high}. Complex graphs for optimization may contain topological structures resistant to simple local perturbations. In the XY regime, these often arise as domain boundaries or vortices. For domain boundaries, the transformation from the excited state to the ground state requires an entire domain to reverse its chirality, representing a significant energy barrier to overcome. Therefore, in this case, local perturbations are not enough to bridge local and global minima.

\section{Graphs for XY Minimization}
\label{Section:Graphs}
To rigorously test CoVeGA’s capabilities, we select graph types that, due to their complex topologies and high energy barriers, present distinct challenges for gradient-based methods.
We consider network graph structures given by the coupling matrix $\mathbf{J}$, resulting in nontrivial minimization instances for simple gradient-based solvers. To do this, we represent the gradient dynamics by  the Kuramoto model (\ref{Kuramoto Equation}), as it represents a gradient descent with respect to XY phases $\theta_{i}$. Moreover, by using the Kuramoto model, we can infer the sizes of the basins of attraction for each locally stable state. This is because, under gradient descent, the system evolves to the closest minima.

\textbf{4-Regular M\"obius Ladder}

\begin{figure}[t]
\centering
     \includegraphics[width=\columnwidth]{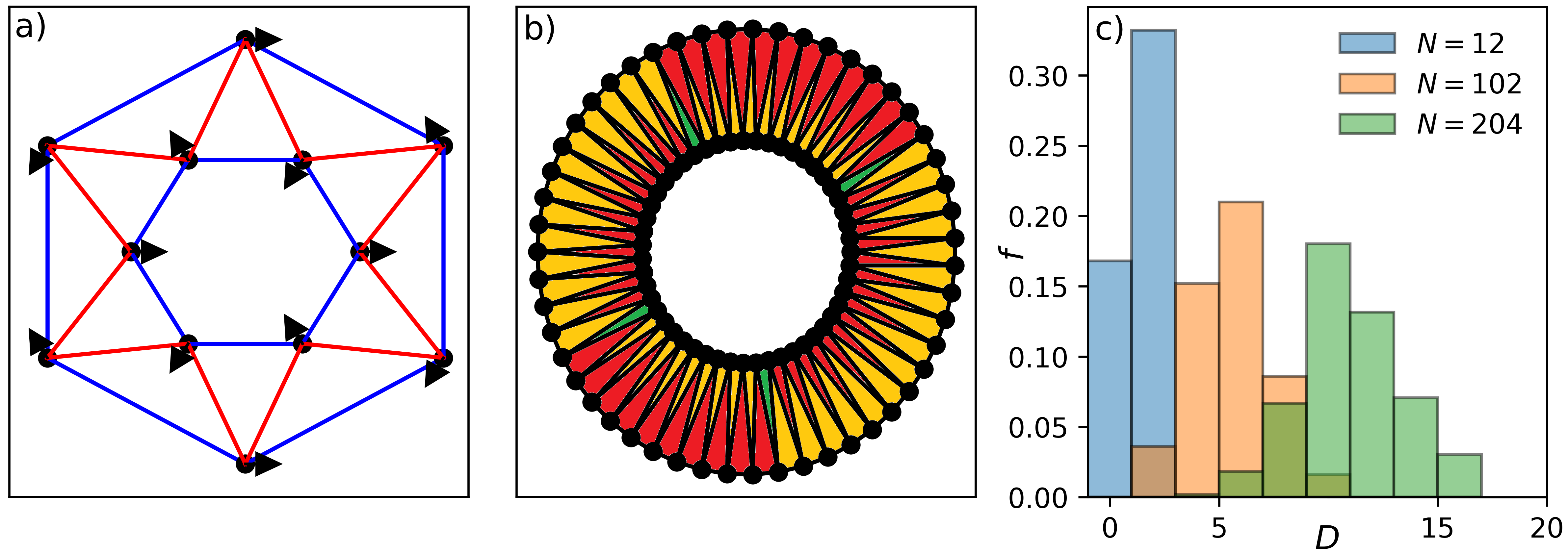}
     \caption{(a) Ground state solution, up to a global phase change, for an $N = 12$ 4-regular M\"obius ladder graph. In this ground state, every XY spin has a phase difference of $\pm 2 \pi / 3$ with each of its nearest neighbors. (b) Excited state of an $N = 102$ 4-regular M\"obius ladder graph with $D = 4$. Here, the phases are not illustrated, and instead color represents the chilarity of each triangular base. For a base with indices $\{ i, j, k\}$, then $C_{ijk} = - 1$ is shown in yellow, $C_{ijk} =  1$ as red, and $C_{ijk} = 0$ as green. (c) Frequency density histogram of excitation parameter $D$ for $1000$ runs of the Kuramoto model on $N = \{12, 102, 204 \}$ 4-regular M\"obius ladder graphs. In each run, initial phases $\theta_{i} (0)$ are chosen uniformly at random from range $[ - \pi, \pi )$, and Eq.~(\ref{Kuramoto Equation}) is solved using the Euler scheme with fixed time step $\Delta t = 0.1$.}
    \label{4Reg Mobius Graph}
\end{figure}

The 4-regular M\"obius ladder network is an unweighted 4-regular circulant graph with antiferromagnetic couplings $J_{ij} \in \{-1, 0\}$ between its $N$ vertices. It is inspired by its 3-regular version, often used as an Ising solver benchmark \cite{kalinin2020complexity, cummins2023classical, cummins2024vector}; see Appendices \ref{Section:Mobius} and \ref{Section:Circulant}. We only consider cases for which $N / 2$ is divisible by $3$, in which case the ground state has no frustrations, and any connected spin pair has phase difference $\pm 2 \pi / 3$. Then, the graph, illustrated in Figs.~(\ref{4Reg Mobius Graph})(a) and (b) consist of $N$ triangular bases connected to each other, with the first and last bases joined, creating a periodic circular geometry. For any triangle with nodes $i,j,k$, we define the chirality $C_{ijk} \in \{ 0,\pm 1 \}$ depending on whether there is a phase winding around these nodes and in which direction. The ground state corresponds to alternating chirality between adjacent triangles. Therefore, we define the excitation parameter $D$ for any 4-regular M\"obius ladder graph as
\begin{equation} \label{Excitation Parameter}
    D = \sum_{ \{ ijk \} \in \mathcal{T}} \left( 1 - | C_{ijk} | \right),
\end{equation}
to account for excitations that occur as domain boundaries, where $\mathcal{T}$ is the set of vertices for each triangle. This is where triangular bases with zero chirality separate domains with triangular bases of alternating chirality $\pm 1$.

The frequency density histogram of excitation parameter $D$ for states recovered by the Kuramoto model (\ref{Kuramoto Equation}) is shown in Fig.~(\ref{4Reg Mobius Graph})(c). Since the system descends to the closest minimum to the initial state according to Eq.~(\ref{Kuramoto Equation}), the histogram represents the volumes of the basins of attraction for each minima. Figure (\ref{4Reg Mobius Graph})(c) shows that as the system size $N$ increases, the volume of the basin of attraction of the ground state decreases as a proportion of the total volume of all basins, and recovering the ground state using standard gradient descent becomes hard.

\textbf{Triangular Lattice}

\begin{figure}[ht]
\centering
     \includegraphics[width=\columnwidth]{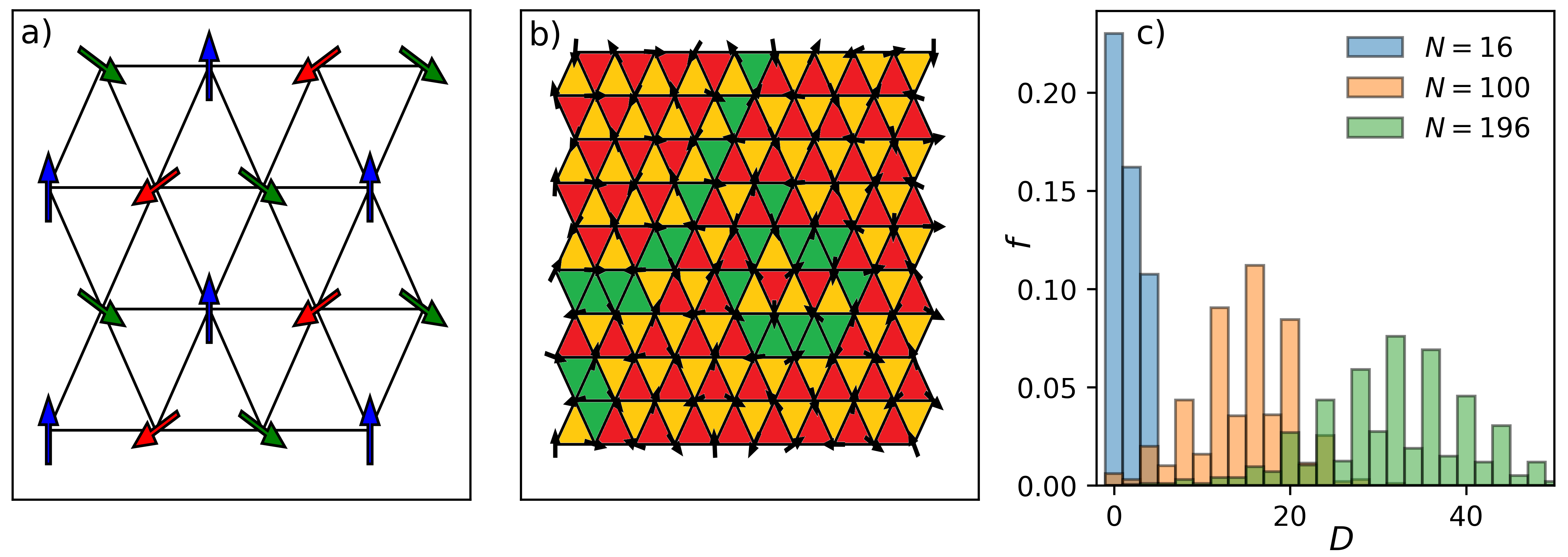}
     \caption{(a) A ground state solution for an $N = 4 \times 4$ triangular lattice graph, where every XY spin has a phase difference of $\pm 2 \pi / 3$ with each of its neighbors. (b) Excited state of an $N = 10 \times 10$ triangular lattice graph. Here, the phases are not illustrated, and instead color represents the chirality of each triangular base. For a base with indices $\{ i, j, k \}$, then $C_{ijk} = - 1$ is show in yellow, $C_{ijk} = + 1$ as red, and $C_{ijk} = 0$ as green. Domains are more likely to exist when part of their boundaries are the edges of the triangular lattice since, on these edges, there are no excited (green) triangular bases. This is a consequence of the graph's non-periodic boundary conditions. (c) Frequency density histogram of excitation parameter $D$ for $1000$ runs of the Kuramoto model on $N = \{16, 100, 196 \}$ triangular lattice graphs. In each run, initial phases $\theta_i (0)$ are chosen uniformly at random from range $[-\pi, \pi)$, and Eq.~(\ref{Kuramoto Equation}) is solved using the Euler scheme with fixed time step $\Delta t = 0.1$.}
    \label{Triangular Lattice}
\end{figure}

The triangular lattice is a  two-dimensional graph consisting of triangles in a regular arrangement. Each of the $N$ vertices corresponds to an XY spin, while edges represent unweighted antiferromagnetic couplings $J_{ij} = -1$. The ground state is recognized by its arrangement of phases, with every XY spin having phase difference $\pm 2 \pi / 3$ with each of its nearest neighbors. Similar to 4-regular M\"obius ladder graphs, we can use the same excitation parameter $D$ from Eq.~(\ref{Excitation Parameter}) to categorize ground and excited states. The frequency density histogram of excitation parameter $D$ for states recovered by the Kuramoto model (\ref{Kuramoto Equation}) is shown in Fig.~(\ref{Triangular Lattice})(c). As the system size $N$ increases, the number of trials in the histogram bin corresponding to the ground state $D = 0$ decreases.

\textbf{Basic Kuratowskian}

Basic Kuratowskian graphs are non-planar graphs that serve as fundamental structures in graph theory; they are sub-graphs of every two-dimensional or three-dimensional non-planar graph. For edges with random weights ${-1, 0, +1}$, the Ising problem on basic Kuratowskian graphs is known to be NP-complete \cite{istrail2000statistical}; however, the analysis of these graphs in the continuous XY spin regime has not been extensively studied. Unlike 4-regular Möbius ladder and triangular lattice graphs, the ground state configuration and energy are not known a priori due to the random coupling weights.

\begin{figure}[ht]
\centering
     \includegraphics[width=\columnwidth]{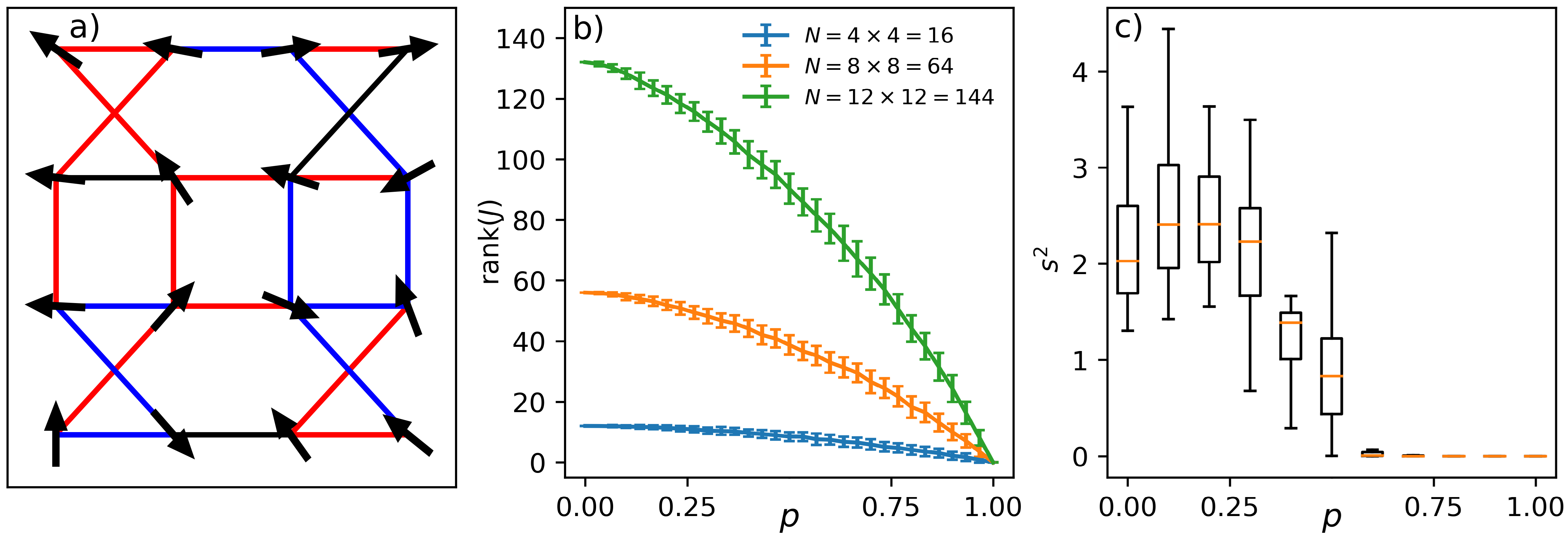}
     \caption{(a) A configuration of XY phases on an $N = 4 \times 4$ basic Kuratowskian graph with $p = 0.2$. Ferromagnetic, antiferromagnetic, and zero couplings are illustrated as red, blue, and black lines, respectively.
     (b) Rank of coupling matrix $\mathbf{J}$ as a function of the probability parameter $p$. Each error bar in (b) corresponding to a different value of $p$ is constructed from 100 Kuratowskian graphs, each with random coupling weights generated by Eq.~(\ref{Kurat pmf}). (c) Box plot distributions of sample variance values $s^2$ from final state energies of Eq.~(\ref{Kuramoto Equation}), obtained from $N = 8 \times 8$ basic Kuratowskian graphs. Each box plot in (c) is obtained by sampling the final state energies over 100 runs on each of the 100 Kuratowskian graphs generated for each value of $p$.}
    \label{Kuratowskian Graph}
\end{figure}

To construct the coupling matrix $\mathbf{J}$ for a basic Kuratowskian graph, we assign each edge in Fig.~(\ref{Kuratowskian Graph})(a) a weight randomly chosen from $W = \{ -1, 0, +1 \}$. We sample the weights $w$ from the discrete probability mass function $p_{W}: W \rightarrow [0, 1]$, where
\begin{equation} \label{Kurat pmf}
    p_{W} (w) = \begin{cases}
        \frac{1}{2} (1 - p) & \text{if } w = -1,\\
        p & \text{if } w = 0,\\
        \frac{1}{2} (1 - p) & \text{if } w = +1.\\
    \end{cases}
\end{equation}
Here, the parameter $0 \leq p \leq 1$ in Eq.~(\ref{Kurat pmf}) controls the rank of $\mathbf{J}$ and influences the hardness of the XY minimization problem, as shown in Figs.~(\ref{Kuratowskian Graph})(b) and (c). As $p$ increases, the number of nonzero entries in $\mathbf{J}$ decreases, and consequently, the rank decreases. In spatial photonic XY and Ising machines, the rank of the coupling matrix is an important feature of the spin network; indeed, some current implementations are only feasible for low-rank coupling matrices \cite{wang2024efficient}.

\section{Results}
\label{Section:Results}

In this section, we compare CoVeGA to the one-dimensional spin version, whose Hamiltonian is given by Eq.~(\ref{Hamiltonian Eq}) and governed by the canonical Stuart-Landau network Eq.~(\ref{Gain-Diss Eq}), as well as to Kuramoto gradient descent and other methods. The phase annealing protocol of CoVeGA begins with $P(0) = 0$ and gradually increases for $t > 0$, ensuring that the local phase difference in each dimension of the complex-field vector $\mathbf{\Psi}_i$ becomes the same. We increase $P(t)$ to a sufficiently large value, so that $\theta{ij}^{(1)} = \theta_{ij}^{(2)}$ holds for every pair $(i, j)$, after which we choose one dimension from which to extract the XY phases. Under this constraint, each $\theta_{i}^{(1)}$ is equivalent to $\theta_{i}^{(2)}$ up to a global phase shift, and hence the choice of dimension is arbitrary. For simplicity, we choose a monotonically increasing function $P(t)$ that has a linear dependence on time, such that $P(t) = \beta t$, although other options may be chosen.

\begin{figure}[t]
\centering
     \includegraphics[width=\columnwidth]{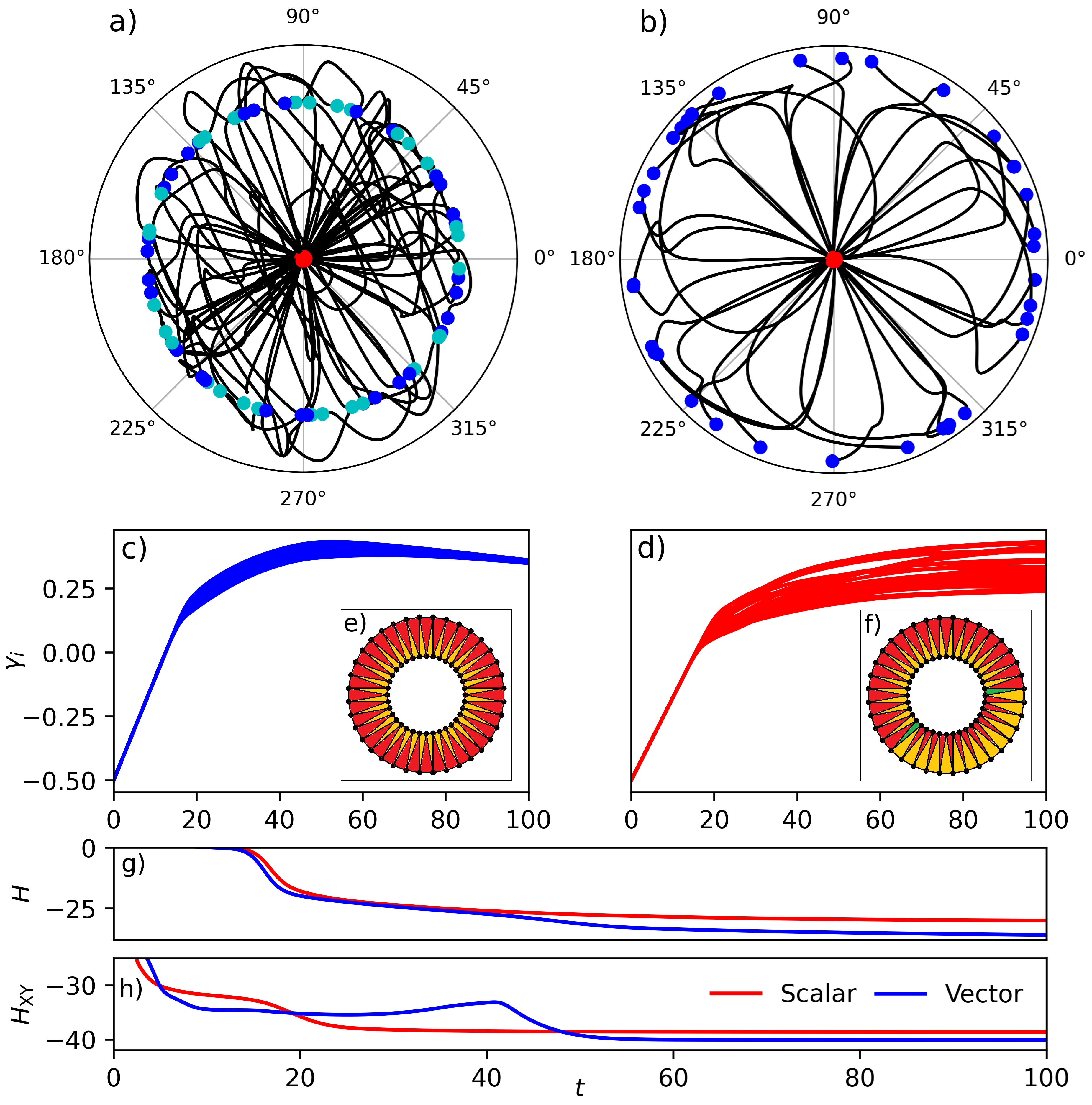}
     \caption{Panels (a) and (b) compare the trajectories of XY spins in the CoVeGA model and the single-dimension Stuart-Landau model, respectively,  for an $N = 36$ 4-regular M\"obius ladder graph. Red circles indicate initial states, while in (a), blue and cyan circles correspond to final states for $\psi_{i}^{(1)}$ and $\psi_{i}^{(2)}$. Final states $\psi_{i}$ in the scalar version (b) are highlighted with blue circles. Panels (c) and (d) illustrate the effective gain $\gamma_{i}$ as the systems evolve. CoVeGA successfully recovers the ground state without frustrations $D = 0$ as seen in inset (e), while the one-dimensional scalar version reaches the excited state $D = 2$ shown in inset (f). Both systems start from equivalent initial conditions, with the one-dimensional scalar version beginning at $\psi_{i} (0) = 0.01 \exp (i a)$ and CoVeGA at ${\bf \Psi}_{i} (0) = 0.01 ( \exp (i a), 0.1 \exp (i b))$, where $a$ and $b$ are uniformly chosen from range $[-\pi, \pi)$. Panel (g) compares the CoVeGA Hamiltonian $H$ against the one-dimensional model Eq.~(\ref{Hamiltonian Eq}), while panel (h) shows the values of the XY Hamiltonians $H_{\rm XY}$.}
    \label{Comparison}
\end{figure}

Figure~(\ref{Comparison}) compares CoVeGA with the one-dimensional Stuart-Landau model of Eq.~(\ref{Hamiltonian Eq}) under equivalent starting conditions on a 4-regular Möbius ladder graph with $N = 36$. The figure demonstrates that while the scalar version fails to find the ground state ($D = 0$), CoVeGA leverages multidimensional phase dynamics to recover the global minimum by bridging minima unreachable by the one-dimensional approach. Additionally, the CoVeGA mechanism allows for phase changes at lower energy costs than required by the scalar version, due to its ability to navigate through the multidimensional space to find the most energy-efficient path to the global minimum.

\begin{figure}[t]
\centering
     \includegraphics[width=\columnwidth]{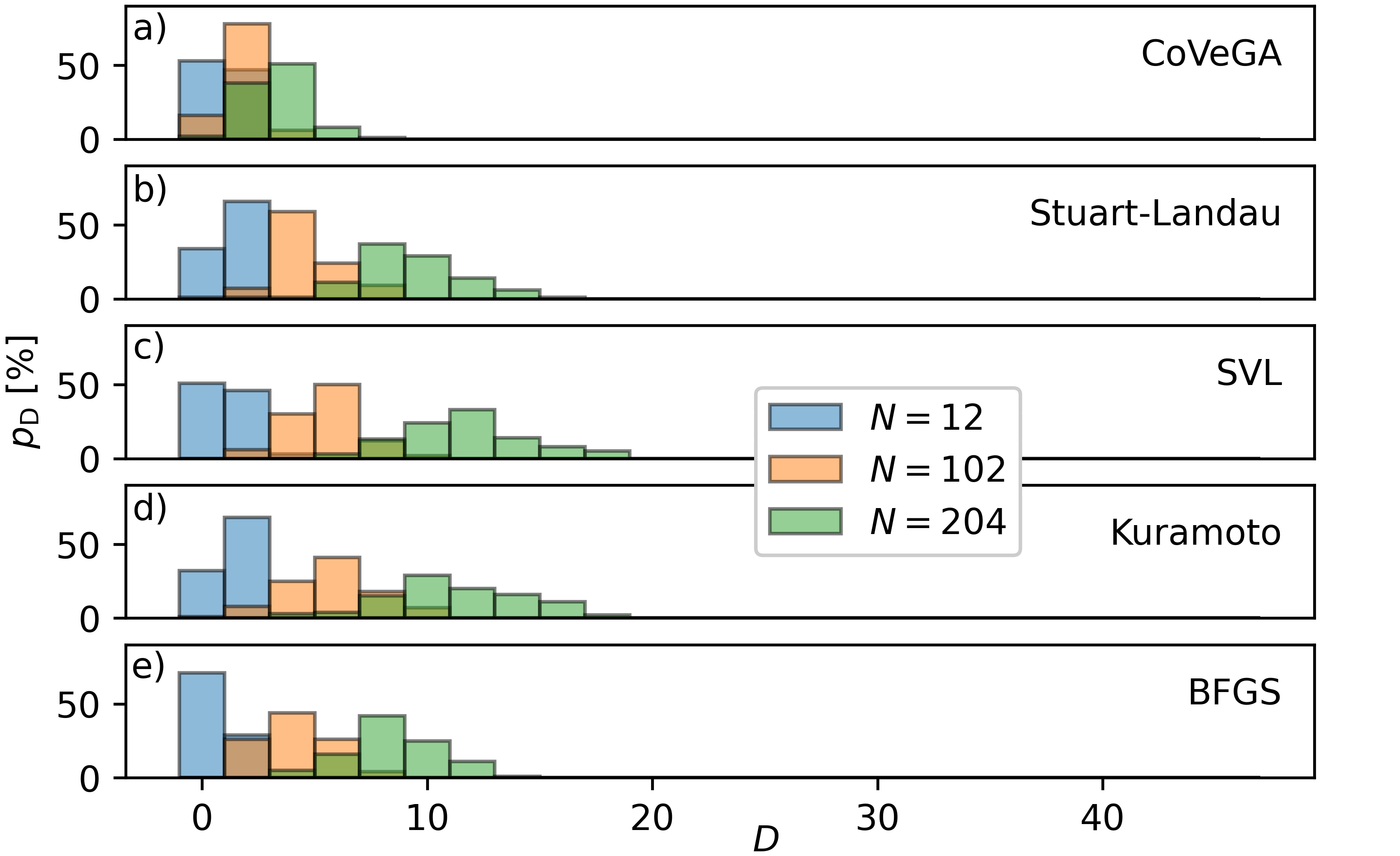}
     \caption{The probability distribution of recovering a state with excitation parameter $D$ for CoVeGA, one-dimensional Stuart-Landau, SVL, Kuramoto, and BFGS methods on $N = 12$ (blue), $102$ (orange), and $204$ (green) 4-regular M\"obius ladder graphs. One thousand runs are used to calculate the probability of recovering excitation parameters $D$ for each value of $N$. Parameter values for CoVeGA are taken from Table (\ref{Optimal_Params}) in Appendix \ref{Section:Materials} \cite{wigley2016fast}, and so are values for the Stuart-Landau network solver where applicable. For SVL, $m = 1.0$, $\xi \sim \mathcal{N} (0, 0.1)$, and $\gamma = 0.9$. In all cases, a fixed time step of $\Delta t = 0.1$ is used with annealing time length $T = 1000$.}
    \label{Probability_Mobius}
\end{figure}

In addition, we compare CoVeGA to the spin-vector Langevin (SVL) model that was proposed as a classical analog of a quantum annealing description using stochastic Langevin time evolution governed by the fluctuation-dissipation theorem \cite{subires2022benchmarking}. SVL is based on the time-dependent Hamiltonian used in quantum annealing $H(t) = A(t) H_0 + B(t) H_{\rm P}$, where the initial Hamiltonian is $H_0 = - \sum_i \sigma_{i}^{x}$. We choose the problem Hamiltonian so that it maps to the XY Hamiltonian as $H_{\rm P} = - \sum_{i, j} J_{ij} (\sigma_{i}^{x} \sigma_{j}^{x} + \sigma_{i}^{z} \sigma_{j}^{z})$, with Pauli operator ${\bf \sigma}_i$ acting on the $i$-th variable. Real annealing functions satisfy boundary conditions $A(0) = B(T) = 1$ and $A(T) = B(0) = 0$, where $T$ is the temporal length of the annealing schedule. If the rate of change of the functions is slow enough, the system stays in the ground state of the instantaneous Hamiltonian so that at $t = T$ the XY Hamiltonian is minimized. The SVL model replaces Pauli operators with real functions of continuous angle $\sigma_{i}^{z} \rightarrow \sin \theta_i$, $\sigma_{i}^{x} \rightarrow \cos \theta_i$, and is therefore a classical annealing Hamiltonian using continuous-valued phases $\theta_i$. SVL dynamics is described by a system of coupled stochastic equations
\begin{equation} \label{SVL_Eq}
    m \Ddot{\theta}_i + b \Dot{\theta}_i + \frac{\partial H (\mathbf{\theta})}{\partial \theta_i} + \xi_i (t) = 0,
\end{equation}
where $m$ is the effective mass, $b$ is the damping constant, and $\xi_i (t)$ is an iid Gaussian noise. For quadratic unconstrained continuous optimization, such as minimizing the XY model, the gradient term in Eq.~(\ref{SVL_Eq}) is
\begin{equation}
    \frac{\partial H (\mathbf{\theta})}{\partial \theta_i} = A(t) \sin \theta_i + B(t) \sum_{j = 1} J_{ij} \sin (\theta_i - \theta_j),
\end{equation}
which in conjunction with fluctuation-dissipation relations $\langle \xi_i (t) \rangle = 0$ and $\langle \xi_i (t) \xi_j (t ') \rangle = \delta_{ij} \delta (t - t ')$, give $2N$ stochastic differential equations: $\md \theta_i = (p_i / m) \md t$ and
\begin{equation}
    \md p_i = \left( \frac{\partial H (\mathbf{\theta})}{\partial \theta_i} + \frac{b}{m} p_i \right) \md t + \md W_i,
\end{equation}
where $\md W_i$ represents a real-valued continuous-time stochastic Wiener process \cite{subires2022benchmarking}.

\begin{figure}[t]
\centering
     \includegraphics[width=\columnwidth]{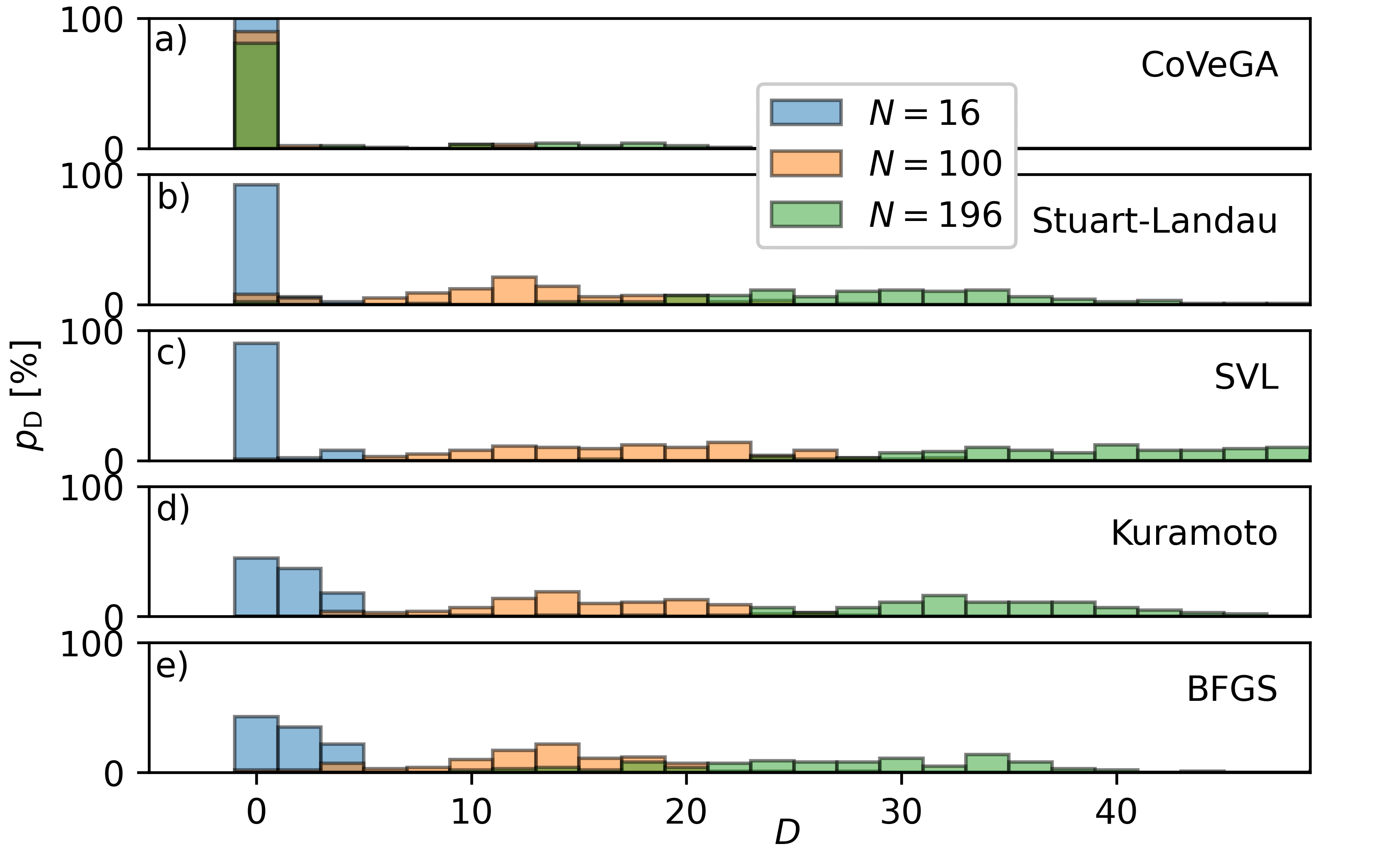}
     \caption{The probability distribution of recovering a state with excitation parameter $D$ for CoVeGA, one-dimensional Stuart-Landau, SVL, Kuramoto, and BFGS methods on $N = 16$ (blue), $100$ (orange), and $194$ (green) triangular lattice graphs. One thousand runs starting from random initial conditions are used to calculate the probability for each method and for each value of $N$.}
    \label{Probability_Triangular}
\end{figure}

CoVeGA distinguishes itself from one-dimensional Stuart-Landau networks, SVL, and the Kuramoto model through its gain-based annealing strategy in higher-dimensional spaces. In Figs.~(\ref{Probability_Mobius}) and (\ref{Probability_Triangular}), we compare CoVeGA to these models and the Broyden-Fletcher-Goldfarb-Shanno (BFGS) algorithm on 4-regular M\"obius ladder graphs and triangular lattice graphs, respectively. BFGS is a classical quasi-Newton method for unconstrained nonlinear optimization that approximates the Hessian matrix to guide the search for a local minimum.

By leveraging a multidimensional approach, the CoVeGA model recovers the ground state ($D = 0$) of 4-regular M\"obius ladder and triangular lattice graphs more often than any of the compared methods. As a result, the distribution of excitations $D$ for CoVeGA is skewed closer to $D = 0$ for all tested values of $N$, compared to the single-dimension Stuart-Landau network, and this effect becomes more pronounced as $N$ increases. While the annealing schedules used in these approaches and in SVL demonstrate efficient convergence to low-energy solutions compared to Kuramoto and BFGS, CoVeGA's higher dimensionality enables XY spins to traverse paths connecting minima and converge to the ground state. This enhances system robustness, making the final system less sensitive to initial conditions. 

Next, we consider basic Kuratowskian graphs, which exhibit significant difficulty for standard gradient descent methods, as illustrated in Fig.~(\ref{Kuratowskian Graph}) for small values of the parameter $p$. However, the randomness of the couplings can introduce statistical convergence issues in small sample sizes. Despite this, graphs with random couplings are popular for benchmarking physical simulators \cite{hamerly2019experimental, harrigan2021quantum, bohm2019poor}.

Because basic Kuratowskian graph instances do not have analytically known ground states, we use the proximity gap metric to evaluate performance. The proximity gap provides a measure of the distance between any solution state and the best-known state by comparing their objective values. More precisely, it is defined as the ratio of the objective value obtained by a given method to the best objective value found among all methods. Here, the objective value is given by the XY Hamiltonian $H_{\rm XY}$. When comparing optimization methods on basic Kuratowskian graphs, we choose $p = 0.1$, the hardest value of $p$, which is the probability that a coupling is set to zero. From Fig.~(\ref{Kuratowskian Graph})(f), we see that this occurs when $p = 0.1$, where the average sample variance $s^{2}$ of Kuramoto runs is maximized.

\begin{figure}[t]
\centering
     \includegraphics[width=\columnwidth]{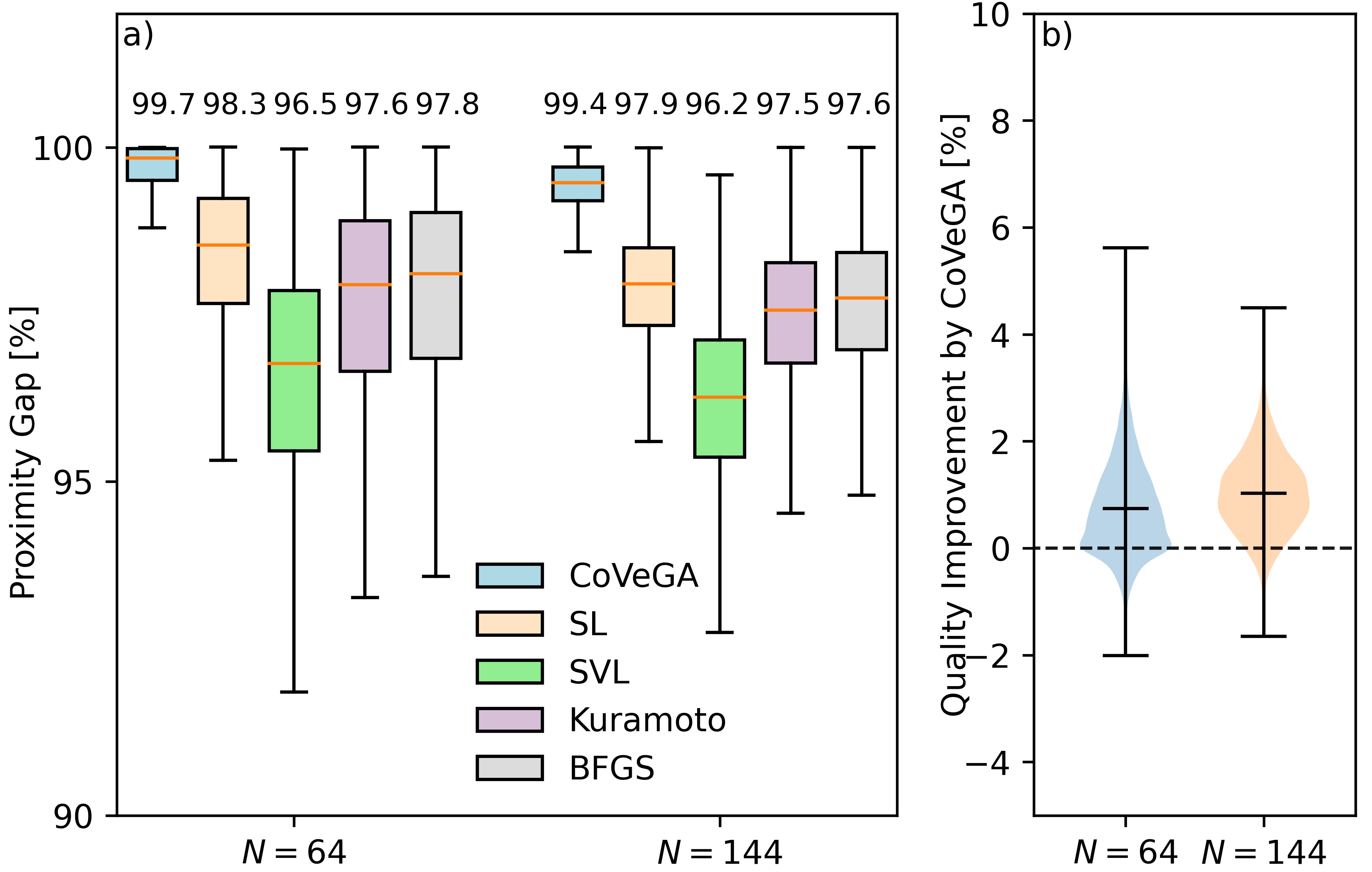}
     \caption{(a) Proximity gap box plots for each method \{CoVeGA, Stuart-Landau (SL), SVL, Kuramoto, BFGS\} on the same $200$ instances of basic Kuratowskian graphs with random bimodel weights, with instances divided equally between graph sizes $N = 64$ and $N = 144$. In all cases, the matrix weight elements are drawn from the discrete probability mass function given by Eq.~(\ref{Kurat pmf}) with $p = 0.1$. (b) Violin plots demonstrate the distribution of CoVeGA's quality improvement performance compared to the best solution found by any competing methods across the $N = 64$ and $N = 144$ basic Kuratowskian graph instances.}
    \label{Proximity}
\end{figure}

Figure~(\ref{Proximity})(a) compares the objective values achieved by CoVeGA with those obtained by one-dimensional Stuart-Landau, SVL, Kuramoto, and BFGS methods for $N = 64$ and $N = 144$ basic Kuratowskian graphs. We define the quality improvement of CoVeGA over another method $X$ in terms of objective values $O$ as $(O_{X} - O_{\rm CoVeGA}) / O_{\rm CoVeGA}$, showcasing these metrics in Fig.~(\ref{Proximity})(b), where $X$ represents the best-performing competing method for each instance.

These results demonstrate that CoVeGA consistently achieves ground states more reliably than one-dimensional models, particularly in larger and more complex systems.
\section{Conclusions}
\label{Section:Conclusions}

 This paper presents the Complex Vector Gain-Based Annealer (CoVeGA). This approach combines multidimensional continuous spin systems, gain-based operations, and soft-amplitude annealing to optimize XY Hamiltonians on challenging graph structures. By exploiting higher-dimensional spaces, CoVeGA improves phase mobility, enabling it to navigate complex energy landscapes more effectively, overcome energy barriers, and avoid local minima. This approach allows CoVeGA to reach optimal solutions with greater efficiency and accuracy compared to one-dimensional methods.

CoVeGA paves the way for the development of continuous analog optimization machines. Applying dimensionality annealing techniques in optical-based spin machines suggests a promising future for ultra-fast computation and accurate ground state recovery, representing a significant advancement in optimization technologies.

\section*{Acknowledgements}

J.S.C.~acknowledges the PhD support from the EPSRC. N.G.B.~acknowledges the support from the HORIZON EIC-2022-PATHFINDERCHALLENGES-01 HEISINGBERG Project 101114978 and Weizmann-UK Make Connection Grant 142568.

\appendix

\section{Complex Ginzburg-Landau Equation}
\label{Section:cGLE}

Gain-based models that describe the dynamics of complex oscillators $\psi_{i}$ are based on physical implementations of laser or non-equilibrium condensate networks coupled to a reservoir $n_{R}$. The systems are described by the generalized complex Ginzburg-Landau equation
\begin{equation} \label{cGLE}
    i \Dot{\psi}_{i} = - \nabla^{2} \psi_{i} + | \psi_{i} |^{2} \psi_{i} + g n_{R} \psi_{i} + i (n_{R} - 1) \psi_{i},
\end{equation}
with reservoir dynamics $\Dot{n}_{R} = - (b_{0} + b_{1} |\psi_{i}|^{2}) n_{R} + P_{i}$, where $g$, $b_{0}$, and $b_{1}$ are dimensionless parameters, and $P_{i}$ is the pumping intensity. Networks of oscillators are constructed by coupling multiple lasers or condensates. In this case, using the tight-binding approximation from the mean-field complex Ginzburg-Landau equation, or the mean-field Maxwell-Bloch equations for laser cavities, Eq.~(\ref{cGLE}) becomes
\begin{equation} \label{Network Eq}
    \Dot{\psi}_{i} = - i |\psi_{i}|^{2} \psi_{i} - \psi_{i} + (1 - i g) \left[ R_{i} \psi_{i} + \alpha \sum_{j \neq i} J_{ij} \psi_{j} \right],
\end{equation}
where $R_{i}$ is the $i$-th reservoir density that follows dynamics $\Dot{R}_{i} = b_{0} (\gamma_{i} - R_{i} - \xi R_{i} |\psi_{i}|^{2})$, where $\xi = b_{1} / b_{0}$, $\gamma_i$ is the effective pumping rate, $\alpha$ quantifies the overall strength of coupling, and $J_{ij}$ is the individual coupling strength between the $i$-th and $j$-th oscillators. In the fast reservoir relaxation limit, $b_{0} \gg 1$ and the reservoir dynamics can be replaced with its steady-state $R_{i} = \gamma_{i} / (1 + \xi |\psi_{i}|^{2})$ to lead to a Stuart-Landau network (\ref{Gain-Diss Eq}) of coupled oscillators. For lasers, the nonlinear self-interactions given by the first term on the right-hand side of Eq.~(\ref{Network Eq}) are zero.

\section{From XY to Ising in Gain-Based Solvers}
\label{Section:XY to Ising}

To obtain the minimum of the Ising Hamiltonian using Eq.~(\ref{Gain-Diss Eq}), a mechanism is required to restrict phases to $\theta_{i} \in \{0, \pi\}$. This may be achieved in many ways, for instance, by (i) coupling the real parts of the field in the coupling term of Eq.~(\ref{Gain-Diss Eq}) as $J_{ij} ( \psi_{j} + \psi_{j}^{*} )$. Then the phases $\theta_{i}$ will automatically be projected onto $0$ or $\pi$ \cite{syed2023physics}. Or, by (ii) externally forcing the system at a frequency resonant with the frequency of the coherent state such that $h(t) \psi^{*(q - 1)}$ is added to the right-hand side of Eq.~(\ref{Gain-Diss Eq}), where $h(t)$ is the real time-dependent magnitude of the resonant pumping term. The presence of this external forcing produces $q$ solutions whose phases differ by $2 \pi / q$ \cite{kalinin2018global}. For $q > 2$, the network emulates the $q$-state Potts model with phases restricted to discrete values that are multiples of $2 \pi / q$. Minimizing the Ising Hamiltonian, therefore, requires taking $q = 2$. For sufficiently large $h(t)$, a feasible Ising solution is obtained by penalizing phases that are not equal to $0$ or $\pi$.

\section{M\"obius Ladder Graphs for XY Networks}
\label{Section:Mobius}

\begin{figure}[t]
\centering
     \includegraphics[width=\columnwidth]{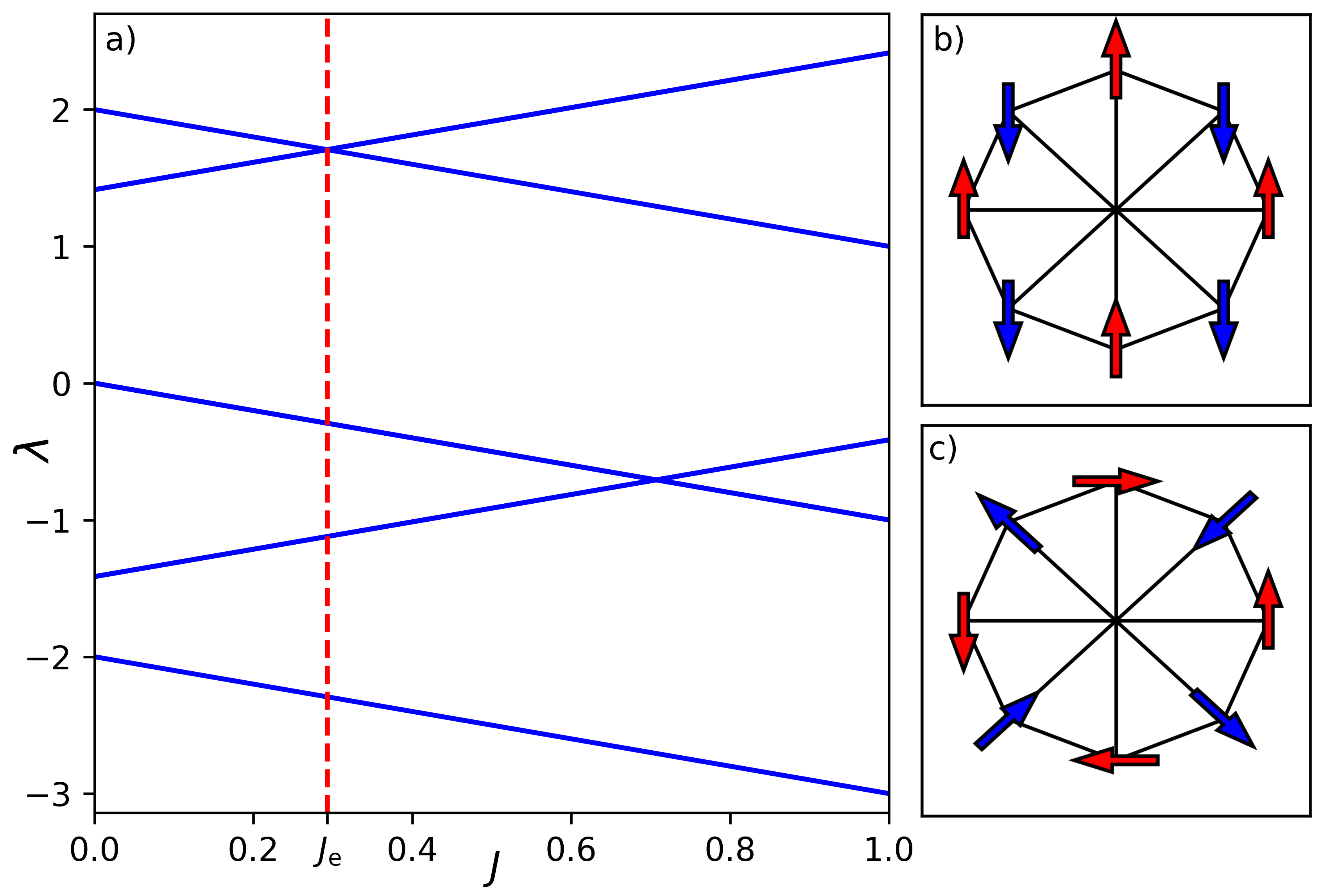}
     \caption{(a) Eigenvalues of an $N = 8$ M\"obius ladder graph as a function of $J$. $J_{\rm e}$, illustrated by the vertical red dashed, is the point at which the two largest eigenvalues cross. XY ground states correspond to $S_{\rm XY}^{(0)}$ for $J < J_{\rm crit}$ and $S_{\rm XY}^{(1)}$ for $J > J_{\rm crit}$, where $J_{\rm crit} = J_{\rm e}$ for the XY model. XY spin configurations for (b) $S_{\rm XY}^{(0)}$, and (c) $S_{\rm XY}^{(1)}$.}
    \label{EValues_XY}
\end{figure}

M\"obius ladder graphs are cyclic graphs with an even number of vertices. They can be visualized as nodes arranged in a ring with antiferromagnetic circle couplings $J_{ij} = -1$ between nearest neighbors, and variable cross-circle couplings $J_{ij} = -J$ between diametrically opposite vertices. At some critical value $J = J_{\rm crit}$, the ground state configuration changes. We denote these states as $S_{0}$ and $S_{1}$, respectively. Notably, below and above $J = J_{\rm e} = 1 - \cos (2 \pi / N)$, the leading eigenvalue of the coupling matrix $\mathbf{J}$ changes from $S_{0}$ to $S_{1}$, where for Ising spins $J_{\rm crit} \neq J_{\rm e}$. Ising Hamiltonian minimization on M\"obius ladder graphs is known to be computationally difficult for certain parameter regimes \cite{cummins2023classical}. Specifically, for cross-circle couplings $J_{\rm e} < J < J_{\rm crit}$, many physically inspired optimization algorithms fail to recover the ground state with high probability \cite{cummins2024vector}. This originates from the behaviour of ${\bf J}$, where for $J_{\rm e} < J < J_{\rm crit}$, the eigenvector of the principal eigenvalue does not correspond to the Ising ground state. Figure~(\ref{EValues_XY})(a) illustrates how these eigenvalues vary as a function of $J$.

Similar to the Ising model, in the continuous XY regime, there are two distinct ground states as $J$ is varied; namely $S_{\rm XY}^{(0)}$ for $J < J_{\rm crit}$ and $S_{\rm XY}^{(1)}$ for $J > J_{\rm crit}$. These have Hamiltonians $H_{\rm XY}^{(0)} = (J - 2) N /2$ and $H_{\rm XY}^{(1)} = [2 \cos ((N - 2) \pi / N) - J] N /2$ respectively, and are shown in Figs.~(\ref{EValues_XY})(b)-(c). The critical value $J_{\rm crit}$ at which these energies intercept is $J_{\rm crit} = 1 - \cos(2 \pi / N)$. This is exactly equal to $J_{\rm e}$ -- the point at which the leading eigenvalues intercept one another. Therefore in the XY regime, $J_{\rm e} = J_{\rm crit}$, and the eigenvectors of the principal eigenvalues correspond to the ground state solutions overall $J$. We conclude that M\"obius ladder graphs are computationally simple in the XY model. This motivates the use of other graphs for benchmarking XY minimizers. Specifically, in Section \ref{Section:Graphs} we utilize frustration and the promotion of domain boundaries to devise suitably hard graphs.

\section{Circulant Graphs}
\label{Section:Circulant}

The weighted adjacency matrix $\mathbf{J}$ of a 4-regular M\"obius ladder graph is circulant because it is constructed through cyclical permutations of any $N$-vector. The graph inherently has vertex permutation symmetry, signifying boundary periodicity and uniform neighborhoods. Any circulant matrix can be expressed as a polynomial of a shift matrix $\mathbf{P}$. For 4-regular M\"obius ladder graphs
\begin{equation}
    \mathbf{J} = - \mathbf{P} - \mathbf{P}^{2} - \mathbf{P}^{N - 2} - \mathbf{P}^{N - 1},
\end{equation}
where $\mathbf{P}$ is the $N \times N$ canonical shift matrix $\mathbf{P} = \bigl( \begin{smallmatrix} \mathbf{0} & \mathbf{I} \\ 1 & \mathbf{0} \end{smallmatrix}\bigr)$, and here $\mathbf{I}$ is an $(N - 1) \times (N - 1)$ identity matrix. The structure of a circulant matrix is contained in any row, and its eigenvalues and eigenvectors can be analytically derived using the $N$ roots of unity of a polynomial $\omega = \exp (2 \pi i / N)$, where the row components $\mathbf{c}$ of the matrix $\mathbf{J}$ act as coefficients. The eigenvectors of $\mathbf{J}$ are the same as the shift matrix $\mathbf{P}$ while the $N$ eigenvalues $\lambda_{n}$ are the components of the product $\mathbf{F} \mathbf{c}$, where $\mathbf{F}$ is the $N \times N$ Fourier matrix. This gives the eigenvalues as
\begin{equation}
    \begin{split}
    \lambda_{n} & = - \omega^{n} - \omega^{2n} - \omega^{(N - 2)n} - \omega^{(N - 1)n} \\
    & = - 4 \cos \left( 2 \pi n / N \right) \cos \left ( 3 \pi n / N \right),
    \end{split}
\end{equation}
for $n = 0, 1, \ldots, N - 1$. The two indices $n$ corresponding to the largest positive eigenvalues are given by
\begin{equation} \label{Max Eigenvalue}
    n = \frac{3N \pm N \pm 6}{6} \pm \left \lfloor \frac{N - 12}{24} \right \rfloor,
\end{equation}
where $N / 2$ is divisible by $3$ and hence $N$ is a factor of $6$. The eigenvectors of $\mathbf{J}$ form an orthogonal basis, and since $\lambda_n \in \mathbb{R}$ and $\mathbf{J} \in \mathbb{R}^{N \times N}$, then we can always choose the eigenvectors to be real. In this case, the eigenvectors corresponding to the eigenvalues with index $n$ are
\begin{equation}
    \mathbf{q}_{n} = \frac{1}{2} \begin{pmatrix}
    2\\
    \omega^{n} + \omega^{-n}\\
    \omega^{2n} + \omega^{-2n}\\
    \vdots\\
    \omega^{(N - 1)n} + \omega^{-(N - 1)n}
    \end{pmatrix} = \begin{pmatrix}
        1\\
        \cos ( \frac{2 \pi n}{N} )\\
        \cos ( \frac{4 \pi n}{N} )\\
        \vdots\\
        \cos ( \frac{2(N - 1) \pi n}{N} )
    \end{pmatrix},
\end{equation}
where the eigenvectors corresponding to the largest eigenvalues are $4$-cyclic vectors $(1, 0, -1, 0, 1, 0, \ldots, -1, 0)$. Indeed, this is the ground state solution of the 4-regular M\"obius ladder graph when you associate each component of the eigenvector with the cosine of the phase. Graphs with circulant coupling matrices are realizable in current experimental platforms \cite{ohadi2016nontrivial, strinati2021all, ayoub2021high}. This, combined with the accessibility to the analytical energy spectrum they allow, makes them natural platforms for analyzing and contrasting properties of different platforms.

\section{Materials and Methods}
\label{Section:Materials}

In all comparisons between methods presented in this paper, numerical integration is performed by the fourth-order Runge-Kutta scheme with discrete time step $\Delta t = 0.1$. For CoVeGA, a linear annealing schedule is chosen such that $P(t) = \beta t$. Each time evolution is to a maximum time $T = 1000$. The total parameter space for CoVeGA is $\{ \varepsilon, \alpha, \beta, \gamma (0) \}$, with optimal choices deduced by the Python machine-learner online optimization package (\texttt{M-LOOP}) \cite{wigley2016fast}. The hyperparameter $\alpha$ is scaled by the inverse of the largest positive eigenvalue of coupling matrix $\mathbf{J}$ such that $\alpha = \alpha^{\prime} / \lambda_{\rm max}$, and the initial effective gain is $\gamma_i (0) = \gamma (0)$ for all $i$. For a selection of graph sizes for 4-regular M\"obius ladder graphs, triangular lattice graphs, and basic Kuratowskian graphs, the optimal parameter choices are detailed in Table (\ref{Optimal_Params}). Since the value of $\lambda_{\rm max}$ varies for random instances of basic Kuratowskian graphs, the values of $\alpha^{\prime}$ in Table (\ref{Optimal_Params}) for this case are chosen as the best over a sample of 100 $p = 0.1$ graph instances.

For results obtained using the one-dimensional Stuart-Landau network given by Eq.~(\ref{Gain-Diss Eq}), parameters $\varepsilon$, $\alpha$, and $\gamma (0)$ are taken from Table (\ref{Optimal_Params}). For SVL we fix $m = 1.0$, $b = 0.9$, and the Gaussian noise $\xi$ is sampled from $N (0, 0.1)$. The Kuramoto model is implemented without hyperparameters, given by Eq.~(\ref{Kuramoto Equation}) with $\alpha = 1$. The BFGS algorithm is ran with the \texttt{SciPy} library using the \texttt{scipy.optimize.minimize} function.

\begin{table}[H]
\centering
\begin{tabular}{ c c c c c c }
 \hline
 \textbf{Graph Type} & $\mathbf{N}$ & $\bm{\varepsilon}$ & $\bm{\alpha^{\prime}}$ & $\bm{\beta}$ & $\bm{\gamma (0)}$ \\
 \hline
 \multirow{3}{*}{~4-Regular M\"obius~} & $12$ & $0.032$ & $2.882$ & $0.020$ & $-0.139$ \\ 
 & $102$ & $0.022$ & $2.402$ & $0.003$ & $-0.453$ \\
 & $204$ & $0.002$ & $2.529$ & $0.008$ & $-1.235$ \\
 \hline
 \multirow{3}{*}{Triangular} & $16$ & $0.045$ & $1.129$ & $0.018$ & $-0.900$ \\ 
 & $100$ & $0.034$ & $2.280$ & $0.004$ & $-0.526$ \\
 & $196$ & $0.010$ & $1.301$ & $0.005$ & $-0.986$ \\
 \hline
 \multirow{3}{*}{Kuratowskian} & $16$ & $0.047$ & $1.148$ & $0.016$ & $-0.295$ \\ 
 & $64$ & $0.023$ & $2.763$ & $0.007$ & $-1.275$ \\
 & $144$ & $0.012$ & $2.745$ & $0.014$ & $-1.285$ \\
 \hline
\end{tabular}
\caption{Optimal sets of hyperparameters for different graphs and sizes.}
\label{Optimal_Params}
\end{table}

\bibliography{references, ReferencesSpacialSpins}

\end{document}